\documentclass[journal,comsoc]{IEEEtran}

\makeatletter

\def\ps@IEEEtitlepagestyle{%
  \def\@oddfoot{\mycopyrightnotice}%
  \def\@evenfoot{}%
}

\def\mycopyrightnotice{%
  {\footnotesize 979-8-3503-9591-4/24/\$31.00~\copyright~2024 IEEE\hfill}
  \gdef\mycopyrightnotice{}
}

\IEEEoverridecommandlockouts
\usepackage{cite}
\usepackage{amsmath,amssymb,amsfonts}
\usepackage{algorithmic}
\usepackage{graphicx}
\usepackage{textcomp}
\def\BibTeX{{\rm B\kern-.05em{\sc i\kern-.025em b}\kern-.08em
    T\kern-.1667em\lower.7ex\hbox{E}\kern-.125emX}}
\begin{document}

\title{Bidirectional Charging Use Cases: Innovations in E-Mobility and Power-Grid Flexibility\\}

\author{
	\IEEEauthorblockN{
		Shangqing Wang\IEEEauthorrefmark{1},
		Juan A. Cabrera\IEEEauthorrefmark{1},
		and
		Frank H. P. Fitzek\IEEEauthorrefmark{1}\IEEEauthorrefmark{2}\\
	}
	\IEEEauthorblockA{
		\IEEEauthorrefmark{1} Deutsche Telekom Chair of Communication Networks, TU Dresden, Germany\\
		}
	\IEEEauthorblockA{
		\IEEEauthorrefmark{2} Centre for Tactile Internet with Human-in-the-Loop (CeTI)\\
		}
	E-mails: \{name.lastname\}@tu-dresden.de
% 	\IEEEauthorrefmark{1}
}

\maketitle

\begin{abstract}
This paper explores the potential of Vehicle-to-Everything (V2X) technology to enhance grid stability and support sustainable mobility in Dresden's Ostra district. By enabling electric vehicles to serve as mobile energy storage units, V2X offers grid stabilization and new business opportunities. We examine pilot projects and business use cases, focusing on Building Integrated Vehicle Energy Solutions (BIVES) and Resilient Energy Storage and Backup (RESB) as stepping stones towards full Vehicle-to-Grid (V2G) implementation. Our analysis highlights the feasibility, advantages, and challenges of implementing V2X in urban settings, underscoring its significant role in transitioning to a resilient, low-carbon urban energy system. The paper concludes with recommendations for addressing technical, regulatory, and business model challenges to accelerate V2X adoption in Dresden and beyond.
\end{abstract}

% \begin{IEEEkeywords}
% Bidirectional Charging, V2X, E-Mobility, Smart Grid, Power-Grid Flexibility, Sustainability Business Models, Use Cases,  Electric Vehicles (EVs), Urban Mobility, Energy Optimization, Pilot Projects, Renewable Energy Integration, Grid Resilience, Energy Storage, Demand Response, Mobilities for EU, Vehicle-to-Grid (V2G), Vehicle-to-Home (V2H), Vehicle-to-Building (V2B), Vehicle-to-Load (V2L), Smart Cities, Dresden, Ostra District
% \end{IEEEkeywords}

\begin{IEEEkeywords}
Bidirectional Charging, V2X, E-Mobility, Smart Grid, Sustainability Business Models.
\end{IEEEkeywords}

\section{Introduction}
Integrating electric vehicles (EVs) into smart grid infrastructure is crucial for sustainable urban mobility and energy optimization~\cite{casellatowards}. This paper explores how bidirectional charging in Dresden's Ostra district can enhance grid stability, reduce energy consumption, and contribute to smart city goals. By implementing Vehicle-to-Home (V2H) and Vehicle-to-Building (V2B) technologies, we aim to pave the way for Vehicle-to-Grid (V2G) integration, which offers significant benefits for energy optimization and emissions reduction.

\section{Background and Motivation}
Smart grid technologies have enhanced the utility of EVs through Vehicle-to-Everything (V2X) technology, which includes various forms of bidirectional charging. This capability leverages EV batteries as flexible energy storage solutions that provide grid support and backup power~\cite{rehman2023comprehensive}. The concept of bidirectional charging gained prominence after the Great East Japan Earthquake in 2011, highlighting EVs' potential as mobile power sources during emergencies. This event catalyzed the development of protocols for using EV batteries as distributed energy resources in crises~\cite{aki2017demand, toswmlowercost}. Since then, various scenarios under the V2X umbrella have emerged, including V2G, Vehicle-to-Load (V2L), V2H, and Vehicle-to-Vehicle (V2V) applications~\cite{aki2017demand}.

\subsection{European Commission's Climate-Neutral and Smart Cities Mission}
As part of the European Commission's Climate Neutral and Smart Cities Mission, the ``Mobilities for EU'' project~\cite{mobilitiesforEU}, funded by Horizon Europe, aims to test and validate innovative mobility technologies in urban environments. The Deutsche Telekom Chair of Communication Networks at the Technical University Dresden (TUD) is responsible for several pilot projects within the project, focusing on the development and deployment of bidirectional charging technologies.

\section{Research Objectives}
Despite extensive research, a significant gap remains between theoretical possibilities and practical business applications of bidirectional charging. This paper aims to bridge this gap within the framework of the "Mobilities for EU" project in Dresden's Ostra district. The primary objective is to analyze business use cases for bidirectional charging and barriers to its widespread adoption. It seeks to identify potential business models, technical requirements, regulatory frameworks, and infrastructural innovations necessary for successful implementation.

This research specifically focuses on the impact of V2B and V2H technologies as preliminary steps towards realizing V2G scenarios. By addressing these factors, the paper aims to provide an initial roadmap for realizing the practical benefits of bidirectional charging technology in Dresden's urban context, contributing to the city's smart city goals and sustainable mobility plans. Ultimately, this work serves as a conceptual exploration of how bidirectional charging can contribute to energy management systems by reducing peak demand, increasing renewable energy utilization, and ultimately leading to CO2 reduction.

\section{Pilot Project Descriptions}
To achieve the objectives of the ``Mobilities for EU'' project, which aims to demonstrate innovative urban mobility solutions and accelerate the transformation of the urban transportation sector, two key pilot projects are being conducted by TU Dresden in the Ostra district~\cite{mobilitiesforEU}:

\subsection{Design and Test of Tuneable/Configurable E-Car for Mobility of People}
This pilot focuses on developing electric vehicles (EVs) with bidirectional charging capabilities, specifically targeting V2H/V2B applications. V2H/V2B systems allow EVs to supply energy directly to homes, enhancing energy efficiency and providing backup power during outages. The pilot aims to achieve a 20\% increase in energy efficiency through user-centric design, collect and analyze data on vehicle performance and user behavior, engage local stakeholders to meet diverse mobility needs, and collaborate with industry partners for potential commercialization.

\subsection{Power-grid Flexibility (Demand-Oriented Transport and E-Charging Solution)}
This pilot aims to optimize energy usage and enhance grid stability through advanced bidirectional charging infrastructure, with a focus on V2G applications. V2G systems enable EVs to discharge energy back into the grid, providing grid services such as frequency regulation and peak load management. The pilot involves analyzing charging demand patterns for efficient grid load management, engaging EV users and grid operators, and targeting Technology Readiness Level 7 (TRL7)~\cite{APRE2022, branas2024trl} through real-world validation.

Both pilots, scheduled from 2024 to 2028, will showcase how V2H/V2B and V2G technologies can enhance grid stability, optimize energy use, and promote sustainable urban mobility in Dresden's unique environment. These initiatives provide a comprehensive framework for integrating bidirectional charging into urban settings, laying the groundwork for future V2G integration.

\subsection{Methodology}
Our methodology involves a phased approach to analyze the impacts of V2B/V2G technologies on energy consumption, cost savings, and CO2 emissions. Initially, we will utilize simulated data to model various scenarios and assess how these technologies can contribute to energy management systems. Future steps may involve exploring co-simulation frameworks to validate our findings further.

\section{Overview of Bidirectional Charging}

\begin{figure}[t]
\centering
\includegraphics[width=0.5\textwidth]{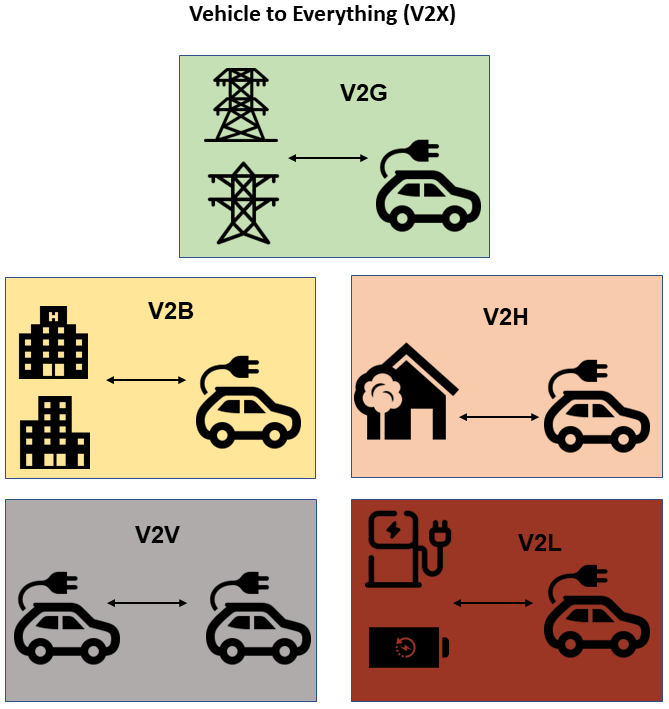}
\caption{Common V2X Bidirectional Charging Scenarios}
\label{fig:V2X}
\end{figure}

Bidirectional charging, a key component of vehicle-to-everything (V2X) technology, enables electric vehicle (EV) batteries to both draw power from and discharge power back to various systems. This technology offers benefits such as grid support, reduced energy consumption, increased renewable energy use, and backup power solutions~\cite{rehman2023comprehensive, adegbohun2024review}. Figure~\ref{fig:V2X} provides a simple illustration of common bidirectional charging scenarios.

\subsection{Types of Common Bidirectional Charging Scenarios}
\subsubsection{V2L}
\textbf{Concept:} Uses EV batteries to power external devices and appliances, providing off-grid electricity and resilience during blackouts~\cite{rehman2023comprehensive}.
\textbf{Target Audience:} Vehicle manufacturers targeting convenience and emergency use.
\subsubsection{V2H/V2B}
\textbf{Concept:} Supplies power to homes or buildings, reducing peak demand and providing backup power. This system enhances energy independence by allowing EVs to supply energy directly to homes, integrating with solar and stationary battery systems~\cite{GSCHWENDTNER2021110977, rehman2023comprehensive}.
\textbf{Target Audience:} Homeowners and businesses seeking energy savings and backup solutions.
\subsubsection{V2V}
\textbf{Concept:} Allows one EV's battery to charge another, useful for fleet management and providing backup power in multi-car scenarios~\cite{rehman2023comprehensive}.
\textbf{Target Audience:} Fleet managers and households with multiple EVs.
\subsubsection{V2G}
\textbf{Concept:} Discharges power from EVs back to the grid, providing grid services and monetizing battery storage capacity. V2G systems support grid stability and can function as virtual power plants (VPPs)~\cite{GSCHWENDTNER2021110977, rehman2023comprehensive}.
\textbf{Target Audience:} Utilities and grid operators.
\subsubsection{Comparison of V2B and V2G}
\textbf{V2B (Vehicle-to-Building)}: Primarily focuses on supplying energy to buildings, enhancing energy efficiency and providing backup power during outages. It allows for localized energy management and supports renewable energy integration at the building level~\cite{borge2021combined}.
\textbf{V2G (Vehicle-to-Grid)}: In contrast, V2G systems enable EVs to discharge energy back to the grid, providing services such as frequency regulation and peak load management. V2G has broader implications for grid stability and energy optimization, allowing for the monetization of stored energy~\cite{GSCHWENDTNER2021110977}.
By understanding these distinctions, stakeholders can better evaluate the potential applications and benefits of bidirectional charging technologies in urban energy systems.

\subsection{Front of the Meter vs. Behind the Meter}
The concepts of "behind the meter" (BTM) and "front of the meter" (FTM) refer to the location and ownership of energy assets relative to the electricity meter~\cite{schmidtke2024multi, rezaeimozafar2022review}. This distinction is crucial for understanding energy management and grid resilience.
\subsubsection{FTM}
FTM assets are located on the utility side of the meter, typically owned by utility companies. They generate and store electricity for distribution, supporting grid stability and capacity~\cite{rezaeimozafar2022review, englberger2021electric, schmidtke2024multi}. In the context of V2G technology, FTM solutions can significantly impact the economy and reduce carbon emissions by providing large-scale energy storage and facilitating renewable energy integration~\cite{KEMPTON2005280}. However, V2G requires greater technological maturity and compatible standards for effective integration compared to BTM solutions~\cite{GSCHWENDTNER2021110977, rehman2023comprehensive}.
\subsubsection{BTM}
BTM assets are situated on the customer side of the meter and are owned by end-users. These systems aim to reduce grid reliance, lower energy costs, improve resilience during outages, and meet sustainability goals~\cite{schmidtke2024multi}. Common BTM solutions include V2L, V2H/V2B, and V2V.
For Dresden's Ostra district, prioritizing BTM solutions offers strategic advantages given current grid infrastructure challenges. These technologies provide flexibility and scalability for efficient energy management while preparing for future FTM asset integration. 

\section{Business Use Cases for Bidirectional Charging}
Bidirectional charging offers numerous potential applications, from grid support to enhancing energy independence~\cite{visakh2022energy, adegbohun2024review}. This section explores seven key business use cases, each demonstrating the technology's versatility and potential impact on Dresden's smart city goals.

\subsection{Grid-Integrated Vehicle Charging (GIVC) Integration (V2G):}
\textbf{Business Case:} Providing grid services and monetizing EV batteries' storage capacity. 
\textbf{End User:} EV owners and fleet operators. 
\textbf{Position:} Front of the Meter.
\textbf{Stakeholder:} Energy utilities, grid operators, EV manufacturers, EV charging network providers, energy aggregators, regulatory authorities.

\subsubsection{Story}
 Sarah participates in Dresden's Grid-integrated Vehicle Charging program. Her EV feeds electricity into the local grid during peak demand, earning her utility bill credits while supporting sustainable mobility goals.

\subsubsection{Obstacle}
Limited grid infrastructure may hinder V2G implementation, as it may not support bidirectional energy flow. Energy providers may hesitate to adopt V2G due to concerns about grid stability and regulatory uncertainties~\cite{adegbohun2024review, gyamfi2013residential, IEA2023Elec}

\subsection{Demand Response and Peak Load Management (DRPLM) (V2G/V2H)}
\textbf{Business Case:} Optimizing energy consumption and costs through flexible EV charging. 
\textbf{End User:} EV owners, residential and commercial customers. 
\textbf{Position:} Front of the Meter.
\textbf{Stakeholders:} Energy utilities, EV manufacturers, charging infrastructure providers, smart grid technology companies, and demand response aggregators. 
\subsubsection{Story} John uses smart charging equipment that adjusts his EV's charge schedule to avoid peak rates, cutting costs and enhancing system reliability in Dresden's Ostra district.

\subsubsection{Obstacle:} Resistance from energy providers to offer dynamic pricing or demand response programs may limit the ability of EV owners to optimize their charging schedules and reduce electricity costs~\cite{perez2017regulatory, IEA2023Elec}.

\subsection{Resilent Energy Storage and Backup (RESB) Solutions (V2H/V2B)}
\textbf{Business Case:} Enhancing grid resilience with EVs as backup power sources.
\textbf{End User:} Homeowners, businesses, critical infrastructure facilities. 
\textbf{Position:} Behind the Meter.
\textbf{Stakeholders:} Energy utilities, microgrid developers, home energy management system providers, renewable energy developers, and emergency response agencies. 
\subsubsection{Story}
The Mobilities for EU project enables the Johnsons to use RESB technology during power outages in a district of Dresden. Their EV provides backup power, ensuring comfort and safety until the grid is restored. This initiative strengthens Dresden's resilience against power disruptions.
\subsubsection{Obstacle}
 Strict regulations may restrict the use of EV batteries for backup power applications due to safety concerns. Energy providers may have policies that limit the use of EVs for backup power, citing grid reliability issues~\cite{IEA2023Elec}.

\subsection{Mobile Energy Solutions for Remote Operations (MESRO) (V2L/V2G)}
\textbf{Business Case:} Portable EVs for remote and temporary power needs.
\textbf{End User:} Event organizers, emergency response teams, construction companies, logistics. 
\textbf{Position:} Behind the Meter
\textbf{Stakeholders:} EV manufacturers, fleet operators, mobile power equipment suppliers, event management companies, and government agencies. 
\subsubsection{Story}
Dresden's Ostra district uses MESRO as part of the Mobilities for EU initiative. During a large outdoor event with limited access to electricity, a fleet of EVs equipped with mobile power supply capabilities ensures a sustainable energy supply. This solution supports Dresden's event infrastructure without being dependent on the grid.
\subsubsection{Obstacle}
 Lack of standardized protocols may challenge seamless integration during remote applications. Energy providers may hesitate to support mobile power supply initiatives due to concerns about grid stability and liability issues~\cite{IEA2023Elec}.

\subsection{Building Integrated Vehicle Energy Solutions (BIVES) (V2H/V2B)}
\textbf{Business Case:} Integrating EVs for building energy independence. 
\textbf{End User:} Homeowners, commercial building owners, tenants. 
\textbf{Position:} Behind the Meter.
\textbf{Stakeholders:} Energy utilities, building energy management system providers, home automation companies, and real estate developers. 
\subsubsection{Story}
Sara participates in the Mobilities for EU project's pilot at the Ostra district sports facility. She charges her EV with solar energy at home and feeds excess power back to the facility's grid. In return, Sara enjoys discounts on facility access, while the sports center benefits from increased energy independence and reduced peak demand, promoting sustainable urban mobility.

\subsubsection{Obstacle}
High upfront costs and integration complexities challenge the widespread adoption of Building Integrated Vehicle Energy Solutions (BIVES) despite its potential long-term benefits~\cite{IEA2023Elec}.

\subsection{Fleet Energy Support Solutions (FESS) (V2G)}
\textbf{Business Case:} Offering grid support services and revenue opportunities for EV fleet operators. 
\textbf{End User:} Fleet operators, energy aggregators. 
\textbf{Position:} Front of the Meter.
\textbf{Stakeholders:} Grid operators, energy market operators, EV manufacturers, fleet management companies. 
\subsubsection{Story}
In Dresden, the Mobilities for EU project supports FESS with bi-directional electric buses. Through a partnership with the local utility company, the transit agency offers grid ancillary services by using the buses' batteries to provide grid frequency regulation and voltage support. This initiative enhances Dresden's urban mobility and energy security.
\subsubsection{Obstacle}
 Regulatory barriers or lack of incentives from energy market operators may impede EV fleet operators' participation in providing grid ancillary services. Energy providers may be cautious about relying on EVs for critical grid support functions due to concerns about reliability and performance~\cite{IEA2023Elec}.

\subsection{Smart Charging and Load Optimization (SCLO) (V2H/V2B)}
\textbf{Business Case:} Optimizing charging infrastructure utilization and improving user experience. 
\textbf{End User:} EV owners, charging station operators. 
\textbf{Position:} Behind the Meter.
\textbf{Stakeholders:} Charging network providers, software developers, smart grid technology vendors, and energy management companies. 
\subsubsection{Story}
Emily, an EV owner, benefits from SCLO under Mobilities for EU in Dresden's Ostra district. Her EV charges optimally, aligning with grid conditions and reducing emissions. The system utilizes real-time data and predictive analytics to optimize charging schedules based on electricity prices, grid demand, and renewable energy availability. This initiative supports Dresden's smart city goals by integrating sustainable transportation and energy management. 
\subsubsection{Obstacle}
The limited availability of smart charging infrastructure and communication protocols may hinder the implementation of SCLO. Energy providers may face challenges coordinating with charging network operators and aggregators to optimize charging schedules and grid operations~\cite{IEA2023Unlock, sumitra2023comprehensive}.

\subsection{The Low-Hanging Fruit: Building Integrated Vehicle Energy Solutions (BIVES) and Resilient Energy Storage and Backup (RESB) as Stepping Stones to V2G} Building Integrated Vehicle Energy Solutions (BIVES) and Resilient Energy Storage and Backup (RESB) represent the most accessible and immediate opportunities for adopting bidirectional charging technologies in Dresden's Ostra district. These V2H/V2B applications offer several advantages that make them ideal stepping stones toward full Vehicle-to-Grid (V2G) implementation:

\subsubsection{Leveraging Existing Infrastructure} BIVES and RESB can be seamlessly integrated into Dresden's Ostra district due to the existing electrical infrastructure. This setup facilitates quicker and more cost-effective integration than V2G applications requiring extensive grid modifications.

\subsubsection{Immediate Benefits to End-Users} BIVES and RESB offer significant benefits to end-users by enhancing both energy independence and resilience. 
\textbf{Energy Independence:}  By charging EVs during peak solar hours and feeding excess energy back into facilities, users can achieve significant energy independence, reducing reliance on the grid and lowering energy costs~\cite{borge2021combined}. \textbf{Carbon Reduction:} V2B strategies can drastically reduce the carbon intensity of buildings, benefiting both employers and employees~\cite{borge2021combined}.
\textbf{Backup Power:} EV owners can use their vehicles as backup power sources during local outages, enhancing Dresden's resilience to power disruptions.

\subsubsection{Grid Support and Stability} 
\textbf{Demand Reduction:} BIVES alleviates pressure on the grid by reducing demand during peak periods, enhancing grid stability. Research indicates that maximum demand during peak hours can be reduced by up to 50\%~\cite{borge2021combined}.
\textbf{Enhanced Resilience:} RESB provides backup power during grid outages, supporting overall grid resilience. 
\textbf{Flexibility and Mobility:} EVs acting as mobile storage units introduce significant flexibility into the energy network without affecting normal building functionality~\cite{borge2021combined}.

\subsubsection{Paving the Way for V2G Adoption}
\textbf{Technical Familiarity:} Implementing BIVES and RESB allows stakeholders to gain experience with bidirectional charging technology on a smaller scale, building the technical expertise necessary for V2G deployment.
\textbf{User Acceptance:} As users become comfortable with V2H/V2B applications, they're more likely to embrace V2G technology in the future.
\textbf{Regulatory Groundwork:} Successful BIVES and RESB implementations can inform policy-making, potentially easing regulatory barriers for V2G adoption.
\textbf{Infrastructure Development: }The charging infrastructure and communication protocols developed for BIVES and RESB can serve as a foundation for future V2G systems.
\textbf{Data Collection:} These initial implementations provide valuable data on energy flows, user behavior, and system performance, informing future V2G strategies.

\section{Case Studies and Real-World Implementations}
Understanding bidirectional charging through real-world examples is crucial for Dresden's implementation strategy. These case studies provide insights into V2X integration challenges and successes in diverse urban settings, offering lessons to enhance deployment in Dresden's Ostra district and beyond.

\subsection{V2G Clarity (UK)} 
The project developed a system for EVs to participate in demand response and provide grid services, demonstrating bidirectional energy flow capabilities. This initiative sets a precedent for future projects by supporting grid stability without compromising battery life~\cite{rehman2023comprehensive}, thereby, promoting sustainable urban mobility and optimizing energy strategies. 

\subsection{eRegio (Germany)}
In this project, eRigio implemented a V2G system integrating EVs, renewable energy sources (RES), and energy storage systems (ESS) to achieve self-consumption, grid stabilization, and peak shaving in regional distribution networks~\cite{rehman2023comprehensive}. This project illustrates effective V2G integration for enhancing grid resilience and supporting sustainable urban energy solutions.

\subsection{Parker (United States)}
Parker project developed and evaluated a V2G system enabling EVs to provide grid services such as spinning reserve, frequency regulation, and reactive power control~\cite{rehman2023comprehensive}. This initiative underscores the feasibility of leveraging EVs to enhance grid reliability and promote sustainable urban mobility solutions. 

\subsection{Nuvve (United States, Europe, Asia)}
The project aimed to commercialize a V2G platform allowing EV owners to sell excess energy stored in batteries to the grid and participate in energy markets. By facilitating energy flexibility and reliability, this project promotes the adoption of EVs and supports sustainable urban energy optimization strategies~\cite{rehman2023comprehensive}.

\subsection{V2G-SYNC (European Union)}
This project developed a V2G system enabling bi-directional energy flow between EVs and the grid, supporting renewable energy integration, and addressing technical, economic, and regulatory challenges~\cite{rehman2023comprehensive}. It demonstrates advanced V2G capabilities crucial for sustainable urban mobility and grid stability. 

\subsection{Nissan}
Nissan has led several V2X initiatives, including the 2012 V2G project demonstrating EVs' grid support capabilities. Ongoing programs include Nissan Energy Perks by EVgo, incentivizing V2G use; Nissan Energy Share, enabling EV-building energy sharing; and Nissan Energy Home, promoting sustainable energy management. Nissan has demonstrated V2G technology in multiple countries, including the US, Japan, Australia, Namibia, and the UK~\cite{rehman2023comprehensive, NissanCommunications}

\subsection{Utrecht's We Drive Solar (Netherlands)}
The We Drive Solar project in Utrecht integrated V2G technology with solar energy, allowing EVs to store and discharge excess power to the grid. It aimed to enhance energy self-sufficiency, support renewables, and improve grid stability. The project demonstrated EVs as flexible energy assets and highlighted the importance of collaboration between municipalities, energy companies, and technology providers for successful V2G integration~\cite{dumiak2022road}.

These case studies showcase diverse applications of bidirectional charging in urban settings. Key lessons for Dresden's Ostra district include developing robust communication systems (V2G Clarity), integrating EVs with renewables (eRegio), using EVs for grid services (Parker), creating incentives for EV owners (Nuvve), and enhancing local energy self-sufficiency (Utrecht). These insights can inform Dresden's implementation strategy, addressing technical challenges, developing business models, and maximizing benefits for EV owners and the grid.

While highlighting bidirectional charging's potential, these studies also reveal common challenges Dresden must address in its implementation strategy, as discussed next.

\section{Challenges and Future Directions}
In the context of the ``Mobilities for EU'' project and its pilot initiatives in Dresden's Ostra district, understanding and addressing the challenges while planning for future advancements is crucial for successfully implementing and scaling bidirectional charging technology.

\subsection{Technical and Infrastructure Challenges: }
\subsubsection{Grid Compatibility and Stability} \textbf{Outdated Grid Infrastructure:} Many urban grids, including those in Dresden, are not yet equipped to handle the dynamic and bidirectional energy flow required for V2G technology. Upgrading these grids to support real-time data exchange and dynamic load management is necessary. \textbf{Integration with Renewable Energy Sources (RES}): Integrating EVs with RES, such as solar and wind energy, requires advanced energy management systems to ensure a balanced and stable energy supply. This integration is essential to enhance the grid's resilience but poses significant technical challenges~\cite{visakh2022energy}.
\subsubsection{Battery Degradation and Efficiency} \textbf{Impact on Battery Life:} Continuous charging and discharging cycles can accelerate battery degradation, affecting the longevity and performance of EVs. Research and development in advanced battery management systems are needed to mitigate these effects. \textbf{Technological Advancements:} Innovations in battery technology, such as enhanced battery management systems, are crucial to support efficient bidirectional charging without compromising battery life~\cite{adegbohun2024review}.
\subsubsection{Advanced Communication Protocols} Effective V2G integration requires sophisticated communication protocols to manage energy flow between EVs and the grid. Developing these protocols is essential to ensure seamless and efficient energy exchange~\cite{sumitra2023comprehensive}.

\subsection{Regulatory and Business Model Challenges}
\subsubsection{Regulatory Frameworks} \textbf{Lack of Standardization:} The absence of standardized regulations across different regions hinders the widespread adoption of V2G technology. Harmonizing these regulations is essential for broader implementation. \textbf{Incentives for Adoption:} Governments need to create incentives for consumers and energy providers to adopt bidirectional charging technologies. These could include tax breaks, subsidies, and funding for infrastructure upgrades~\cite{IEA2023Unlock}.
\subsubsection{Business Models} \textbf{Economic Viability: } Developing viable, robust scenarios with business models that make V2G economically attractive for all stakeholders is challenging. Models must consider costs related to infrastructure upgrades, technology development, and potential savings from energy optimization~\cite{IEA2023Elec, IEA2023Unlock}.

\section{Conclusions and Future Research Directions}
The "Mobilities for EU" project in Dresden's Ostra District is leveraging pilot projects to validate V2G technologies, providing insights into real-world challenges and scalable solutions for urban implementation. By adopting a phased approach that prioritizes V2H/V2B technologies, the project lays the groundwork for future V2G integration as infrastructure, regulations, and public awareness evolves.

Addressing technical, regulatory, and business model challenges is crucial for the successful adoption of bidirectional charging. These technologies can significantly enhance urban grid stability, optimize energy use, and support sustainable mobility in cities like Dresden. Future research should focus on overcoming implementation barriers and quantifying the long-term benefits of V2X technologies in diverse urban settings. Additionally, exploring co-simulation frameworks will help validate findings from pilot projects.

This research contributes to the theoretical understanding of bidirectional charging by demonstrating its potential as a pathway to full V2G integration. By analyzing the impacts of V2H/V2B technologies, we aim to provide a foundational framework that informs future research and practical implementations in urban energy management systems.

\section*{Acknowledgement}

This work is funded by the European Union under Grant Agreement No 101139666, MOBILITIES FOR EU. Views and opinions expressed are those of the authors only and do not necessarily reflect those of the European Union or the European Climate, Infrastructure and Environment Executive Agency (CINEA). Neither the European Union nor the granting authority can be held responsible for them. It is supported by  the Economic Affairs and Climate Action (BMWK) under project ID 03EI6082A, DymoBat, the German Research Foundation (DFG) as part of Germany's Excellence Strategy—EXC 2050/1—Cluster of Excellence “Centre for Tactile Internet with Human-in-the-Loop” (CeTI) of Technische Universität Dresden under project ID 390696704 and the Federal Ministry of Education and Research (BMBF) in the program of “Souverän. Digital. Vernetzt.” Joint project 6G-life, grant number 16KISK001K.

\bibliographystyle{IEEEtran}
\bibliography{IEEEabrv,references}
\end{document}